\documentclass[12pt]{amsart}

\usepackage{caption}

\usepackage{amsmath, amssymb, graphics}
\usepackage{graphicx}
\usepackage{epstopdf}
\usepackage{booktabs}
\usepackage[latin1]{inputenc}
\usepackage{hyperref}
\usepackage{caption}
\hypersetup{colorlinks=true,linktoc=page,linkcolor=blue,citecolor=red,urlcolor=cyan}

\numberwithin{equation}{section}

\title{Thermodynamics in the NC disc}
\author{S.~A.~Franchino-Vi\~nas}
\address{Theoretisch-Physikalisches Institut, Friedrich Schiller Universit\"at Jena, Max Wien Platz 1, 07743 Jena, Germany,  and \newline 
Departamento de F\'isica, Facultad de Ciencias Exactas
Universidad Nacional de La Plata, C.C.\ 67 (1900), La Plata, Argentina\newline}
\email{safranchino@uni-jena.de}
\author{ P.~Pisani}
\address{IFLP-CONICET/ Departamento de F\'isica, Facultad de Ciencias Exactas\newline
Universidad Nacional de La Plata, C.C.\ 67 (1900), La Plata, Argentina\newline}
\email{pisani@fisica.unlp.edu.ar}

\begin{document}


\maketitle

\begin{abstract}
We study the thermodynamics of a scalar field on a noncommutative disc implementing the boundary as the limit case of an interaction with an appropriately chosen confining background. We explicitly obtain expressions for thermodynamic potentials of gases of particles obeying different statistics. In order to do that, we derive an asymptotic expansion for the density of the zeros of Laguerre polynomials. As a result we prove that the Bose-Einstein condensation in the noncommutative disc does not take place.
\end{abstract}
\maketitle

\section{Introduction}

After the seminal papers of Connes and Lott \cite{Connes:1990qp} and thanks to the impulse given by works as those of Doplicher et al.\ \cite{Doplicher:1994zv} and Seiberg and Witten \cite{Seiberg:1999vs}, noncommutative (NC) quantum field theories have become an intense field of research. Indeed, the derivation of the Standard Model from a spectral triple in a noncommutative geometry or the arising NC QFT in a string theory with D-branes in the presence of a Neveu-Schwarz background have been perceived as clues pointing to the existence of noncommutativity. Moreover, the dynamics of quantum fields on NC spaces---first introduced in \cite{Snyder:1946qz}---exhibits many fascinating properties owing to non-locality and to the existence of a minimal area \cite{Douglas:2001ba,Szabo:2001kg, Perivolaropoulos:2017rgq}. If these are to be true, the most probable sources of its confirmation might come from astrophysical data such as the cosmic microwave background radiation, gamma ray bursts or even neutron stars, corresponding all to some of the most extremal regimes observed in nature \cite{AmelinoCamelia:2008qg}.

In this article we study the thermodynamics of a free scalar field on a NC generalization of the two dimensional disc. The choice of a compact base manifold aims at providing a more realistic description of physical settings. However, the subtleties involved in the representation of boundaries in NC space have lead to different non-equivalent formulations. In this article we define the NC disc as an appropriate limit of a confining background in Moyal plane and, in this way, avoid the ambiguities inherent to the implementation of boundary conditions. Previous work on the NC disc has been performed in \cite{Falomir:2013vaa} and other NC manifolds with boundaries have been studied in \cite{Lizzi:2003ru,Lizzi:2005zx,Lizzi:2006bu,Balachandran:2003vm,Lizzi:2012xy,Dias-Prata,Kryukov:2004ab}.

Moyal plane is defined by the algebra $[\hat{x}_1,\hat{x}_2]=2i\theta$, where $\hat{x}_i$ are hermitian operators and $\theta$ is the real parameter which introduces noncommutativity: non-locality and the existence of a minimal area follow straightforwardly from a nonvanishing parameter $\theta$. As a consequence, the usual point-like states do not exist and the definition of a boundary in Moyal plane becomes ambiguous.

In \cite{Falomir:2013vaa} quantum fields have been confined in a disc of radius $R$ as the result of a limiting process in which the fields interact with a rotationally invariant background that, in the limit, becomes infinite at distances larger than $R$. The spectrum of one-loop oscillations has been explicitly determined and some peculiar properties have been exposed.

Those results motivate the analytical study of statistical quantities for a gas composed of free particles. Some interesting results on the thermodynamics of NC gases had been previously obtained in \cite{Houca:2018afh,Vinas:2016jmp,Halder:2016qwz,Brito:2015csa,Scholtz:2008cb,Kriel:2011aa,Scholtz:2015fba}. In \cite{Scholtz:2008cb,Kriel:2011aa,Scholtz:2015fba}, for example, it is shown that the behaviour of some Fermi gases at high densities and low temperatures is altered by noncommutativity, which may give a clue to understand some aspects of the very early universe. However, in this approach an angular momentum parameter is introduced to deal with the existence of states with increasingly large angular momentum. In our approach there is no need of doing so, since the use of a both left- and right-Moyal background---i.e.\ the existence of a parity symmetry in the underlying QFT action---effectively truncates the available angular momenta of the states.

At this point, a remark regarding the perturbative study of noncommutative corrections to physical systems is in order. Several quantum models in Moyal space admit an explicit solution for arbitrary values of the NC parameter $\theta$ but, in general, a perturbative determination of the consequences of noncommutativity are usually not available since these explicit solutions can only be expanded in inverse powers of $\theta$, thus describing the regime known as ``extreme'' noncommutativity. Other studies are based on a perturbative treatment of Moyal interactions but, for not unrelated reasons, this does not capture the full non-local character of noncommutativity. On the contrary, in this article we are able to study the small-$\theta$ expansion of expressions which represent the full solution for any $\theta\in\mathbb{R}$. In this way, we determine the leading NC corrections to physical quantities which intrinsically contain the whole non-local effects of noncommutativity.

This article is organized as follows: first, in Section \ref{s2} we summarize some previous results regarding the NC disc. Then, in Section \ref{s-preliminaries} we state the definition of the thermodynamical quantities that will be employed all through the text. This paves the way to analyze, in Section \ref{s-low}, the leading noncommutative correction to the partition function on the disc. As a first result we derive an asymptotic expansion for the density of Laguerre zeros; this enables us to obtain an expression for the partition function for small values of $\theta$. In this analysis, we consider particles obeying three different statistics: Maxwell-Boltzmann, Fermi-Dirac and Bose-Einstein. One of the main results is then stated: Bose-Einstein Condensation (BEC) does not occur in the NC disc.

In Section \ref{s-high} we compute the partition function of a gas in the high-temperature regime for all three aforementioned  statistics. Then, we consider thermodynamical quantities such as the entropy, the mean energy and the heat capacity at constant volume. In particular, we observe the uncommon behaviour of the specific heat and of the corrections to the equation of state of the ideal gas.

Finally, we present our conclusions in Section \ref{s5}. Additionally, we sketch some extra proofs regarding the asymptotic density of Laguerre zeros in Appendix \ref{contributions_euler_maclaurin} and regarding the BEC in Appendix \ref{appendix.nobec}, while in \ref{appendix.fermion} we include the complete expressions corresponding to several thermodynamic quantities of fermions in the low temperature regime.

\medskip

\section{Scalar field in the noncommutative disc}\label{s2}
In this section we rederive the spectrum of the NC disc following the lines of \cite{Falomir:2013vaa}, where the formal aspects are considered in more detail. Independently of the statistics, we will study massive spinless particles, which we describe by a real scalar field $\phi(t,x)$ on Moyal plane, whose coordinates we denote by $x=(x_1,x_2)$; we consider Minkowski time $t$ as an ordinary commuting parameter. We introduce an interaction with a background $V=V(r^2)$ in Moyal space ($r^2:=x_ix_i$) which after taking an appropriate limit will confine the quantum field in a disc. The algebra satisfied by the coordinate operators,
\begin{align}
    [\hat{x}_1, \hat{x}_2]= 2 i \theta\,,
\end{align}
can be represented as the algebra of ordinary fields with the multiplication replaced by the Moyal product \cite{Groenewold:1946kp,Moyal:1949sk}:
\begin{equation}\label{mp}
  \phi(x)\star \psi(x):=\phi(x)\,e^{i\theta\,\epsilon_{ij}\overleftarrow{\partial}_i\overrightarrow{\partial}_j}\,\psi(x)\, ,
\end{equation}
where $\epsilon_{ij}$ is the Levi-Civita antysimmetric tensor. Since the exponent contains an antisymmetric combination of derivatives, the difference between the Moyal product and the ordinary commutative product can be regarded as a total derivative. Therefore, the Moyal product has no effect on quadratic terms in the action and only modifies interaction terms. Under this representation, the action of the scalar particle in interaction with the confining background reads
\begin{equation}\label{action}
    S[\phi]=\frac12\int_{\mathbb{R}\times\mathbb{R}^2}dtdx\,\left\{(\partial_t \phi)^2-(\partial_x \phi)^2-m^2\,\phi^2-V\star\phi\star\phi\right\}\,.
\end{equation}
Note that the interaction term $V\star\phi\star\phi$ is invariant under cyclic permutations so no ordering ambiguity arises.

To determine the oscillation modes which span the Hilbert space of the quantized field we study the classical equation
\begin{equation}\label{olqf}
    \delta_\phi S=\left\{-\partial^2_t \phi+\partial^2_x \phi-m^2\,\phi-\tfrac12 V\star\phi-\tfrac12 \phi\star V\right\}
    =0\,.
\end{equation}
We remark that the interaction term in \eqref{action} naturally introduces both left- and right-Moyal multiplications in \eqref{olqf}: this ensures parity invariance and leads to a finite-dimensional Hilbert space for the noncommutative quantum field.

The classical solutions $\phi(t,x)=e^{-i\omega t}\phi(x)$, with definite frequency $\omega$, satisfy
\begin{equation}\label{olqf2}
    A\phi(x)=(\omega^2-m^2)\phi(x)\,,
\end{equation}
where
\begin{equation}\label{opa}
    A:=-\partial^2_x+\frac12\, V(x_i^+x_i^+)+\frac12\, V(x_i^-x_i^-)\,.
\end{equation}
The operators $x_i^{\pm}:=x_i\mp i\theta\,\epsilon_{ij}\partial_j$ implement left- and right-Moyal multiplications by the background $V(r^2)$. Next, to confine the field in a disc of radius $R$, we specify the background as
\begin{equation}\label{bg}
    V(r^2):=\frac{2\Lambda}{\theta}\,\Theta(r^2-R^2)\,,
\end{equation}
where the $\Lambda\rightarrow\infty$ limit is implicit; $\Theta(\cdot)$ is the step-function (defined as $1$ if its argument vanishes).

Eqs.\ \eqref{olqf} and \eqref{olqf2} indicate that the Hilbert space of the quantized field is spanned by the eigenstates of the operator $A$. As we mentioned, we will show that these eigenstates span a finite dimensional space. To exploit the rotational invariance, we define the creation and annihilation operators
\begin{equation}\label{creation.operators}
    a_{\pm}:=\frac{1}{2\sqrt{\theta}}(x^\pm_1 \mp i x^\pm_2)\,,
\end{equation}
which satisfy the algebra
\begin{equation}
    [a_+,a^\dagger_+]=[a_-,a^\dagger_-]=1\,,
\end{equation}
and generate the Fock space of quantum states with circular polarization. Correspondingly, the number and angular momentum operators read
\begin{align}
    N_{\pm}=a^\dagger_\pm a_\pm\,,\qquad
    L=N_+-N_-\,.
\end{align}
With these definitions the operator $A$ can be written as
\begin{equation}\label{opan2}
    \theta\, A=N_++N_-+1-a^\dagger_+a^\dagger_--a_+a_-
    +\Lambda\,\left\{ \Theta(N_+-N)+\Theta(N_--N)\right\}\,,
\end{equation}
where $N$ is defined through the ceiling part function $\lceil \cdot \rceil$:
\begin{equation}\label{Ndef}
N:=\lceil R^2/4\theta-1/2 \rceil\,.
\end{equation}
Roughly speaking, this positive integer represents the quotient between the area of the disc, $\pi R^2$, and the fundamental area in noncommutative space, $4\pi\theta$. Consequently $N$ is a measure of the noncommutativity of the disc and we can assume $N\gg 1$ in the almost commutative case, i.e., for $\theta\ll R^2$.

In order to find the spectrum of $A$ as given in expression \eqref{opan2}, we expand its eigenstates in the Fock space basis generated by $a^\dagger_\pm$. Since the operator $A$ is rotationally invariant, it is convenient to use quantum numbers $(\ell,n)$: the angular momentum $\ell$ and a positive integer $n$, denoting the eigenvalues of $L$ and $N_-$, respectively. Due to parity invariance we may consider $\ell\geq 0$.

First, let us explore the consequences of the limit $\Lambda\to\infty$. According to expression \eqref{opan2}, this limit implies that eigenfunctions with finite energy must satisfy $N_\pm<N$, or, equivalently,
\begin{align}
    \ell<N\qquad n<N-\ell\,.\label{cond}
\end{align}
On the other hand, the expansion of an eigenstate of eigenvalue $\lambda$ in terms of the Fock space basis is given by certain coefficients $c^\ell_n(\lambda)$ which satisfy a recursion relation determined by expression \eqref{opan2}. For $\ell<N$ and $n<N-\ell$ the step-functions in \eqref{opan2} vanish and one finds the following solution to this recursion relation:
\begin{equation}\label{cl}
    c^\ell_n(\lambda)=\frac{\sqrt{n!}}{\sqrt{(n+\ell)!}}\,L^\ell_n(\lambda)
    \qquad {\rm for\ }\ell<N\ {\rm and\ }n< N-\ell\,,
\end{equation}
in terms of Laguerre polynomials. Note that the recursion relations allow one to extend the validity of this expression to $c^\ell_n(\lambda)$ with $n=N-\ell$. However, conditions \eqref{cond} impose that the coefficient $c^\ell_{N-\ell}(\lambda)$ vanishes. Thus,
\begin{align}\label{fullspec}
    L^\ell_{N-\ell}(\lambda)=0\,.
\end{align}
This condition, together with parity invariance, determines the spectrum of the NC disc: the eigenvalues $\lambda$ are given by the zeros $\lambda^\ell_k$ of the Laguerre polynomial $L^{|\ell|}_{N-|\ell|}(\lambda)$, with $|\ell|<N$ and $k=1,2,\ldots,N-|\ell|$. This determines an $N^2$-dimensional Hilbert space.

\section{Thermodynamics in the NC disc: preliminaries}\label{s-preliminaries}

As it is known from statistical mechanics, the grand-canonical partition function $\mathcal{Z}$ for fermions ($a=1$), bosons ($a=-1$), and particles described by Maxwell-Boltzmann ($a=0$) statistics is given by a sum over all the allowed energy states. In the case of a gas in the NC disc, the partition function, using the notation $\lambda_k^\ell$ of Section \ref{s2} to denote the $k$-th zero of the Laguerre polynomial $L^{|\ell|}_{N-|\ell|}(\lambda)$, reads\footnote{The classical gas, i.e.\ that obtained considering Maxwell-Boltzmann statistics, is the usual approximation for diluted gases---non-interacting particles--- where $\exp\left\{\beta\left(\frac{\lambda^\ell_k}{\theta}-\mu\right)\right\}\gg 1$ for all the available energy states. This also usually implies that the chemical potential $\mu$ must be negative enough to guarantee the inequality.}
\begin{equation}\label{pfgc}
  \log\mathcal{Z}_{\theta}=\frac{1}{a} \sum_{\ell,k}
  \log\left(1+ a e^{-\beta\left(\frac{\lambda^\ell_k}{\theta}-\mu\right)}\right)\,.
\end{equation}
In this expression the Lagrange multipliers $\beta$ and $\mu$ are the inverse temperature and the chemical potential. Recall that from eq.\ \eqref{pfgc} one can compute every relevant thermodynamic quantity; in particular, the mean particle number $\mathcal{N}$, mean energy $E$,  pressure $p$, entropy $S$ and constant volume heat capacity $C_V$ are given by
\begin{align}
  \begin{split}
  \mathcal{N}&=\frac1\beta\,\partial_\mu\log{\mathcal{Z}_{\theta}}\,,\label{termodinamica}\\[2.5mm]
  E&=-\frac{1}{\beta}\left(\beta\,\partial_\beta-\mu\,\partial_\mu\right)\log{\mathcal{Z}_{\theta}}\,,\\[2.5mm]
  p&=\frac{1}{\pi R^2 \beta}\,\log{\mathcal{Z}_{\theta}}\,,\\[2.5mm]
  S&= \log\mathcal{Z}_{\theta}-\beta\partial_\beta\log\mathcal{Z}_{\theta}\,, \\[2.5mm]
  C_V&=\left(\frac{\partial E}{\partial T}\right)_{\mathcal{N},V}
  =-\beta^2\,\partial_\beta E
  +\beta^2\,\partial_\mu E\,\partial_\beta\mathcal{N}\,\left(\partial_\mu\mathcal{N}\right)^{-1}\,,
\end{split}
\end{align}
where partial derivatives correspond to constant $\beta$, $\mu$ or $V$, unless it is otherwise specified by the subscript notation usual in thermodynamics. These formulae, valid for all statistics, will prove useful in the following analysis.

\section{Large-$N$ approximation}\label{s-low}

We will show in this section how to perform an expansion of the partition function valid for sufficiently small $\theta$. We remark that we will not perform a perturbative approximation of the Moyal interaction, which would spoil the non-local character of noncommutativity, but will consider instead the leading NC contributions to the partition function as computed from the full NC spectrum given by \eqref{fullspec}. The key idea is to expand eq.\ \eqref{pfgc} by noting that the sum over Laguerre polynomial's zeros may be replaced by an integral over their asymptotic density in the large N regime\footnote{This limit implies $R^2\gg \theta$. However, as will be seen next, our expansion also demands that $\beta$ cannot be much smaller than $\theta$ so that for extremely high temperatures---much larger than $1/\theta$---the results of this section cease to hold.}. Therefore, before turning to the analysis of the thermodynamics we briefly work out below such an expression for this density.

\subsection{Asymptotic density of zeros of Laguerre polynomials}\label{appendix.density.zeros}

For simplicity, in this section we will denote by $z_k$, with $k=1,2,\ldots,K$, the zeros of the rescaled Laguerre polynomial $L_K^{cK}(Kz)$, for fixed $c$ and $K$. We will prove that for any adequately smooth function $f(z)$ the following expansion is valid for large $K$
\begin{align}\label{thesum}
 \frac{1}{K}\sum_k f(z_k)\sim \int_{\mathbb{R}} \left(D^{(0)}(c,z)+D^{(1)}(c,z)\frac{1}{K}+\cdots\right)\, f(z)\,  dz\,.
\end{align}
The function $D^{(0)}(c,z)$ is usually called the asymptotic density of the Laguerre zeros. In order to find a closed expression for the functions $D^{(i)}(c,z)$ we will consider the logarithmic derivative of the rescaled Laguerre polynomial, viz.
\begin{align}
  \Psi(z)=\frac{\left(L_K^{cK}\right)'(Kz)}{L_K^{cK}(Kz)}\,.
\end{align}
The sum in the r.h.s.\ of \eqref{thesum} can thus be expressed as an integral in the complex plane over a closed contour $\Gamma$ that encircles counter-clockwise a large enough but fixed real interval\footnote{The zeros of the polynomial $L_K^{cK}(Kz)$ are known to be real and contained in a bounded region.}
\begin{align}
 \frac{1}{K}\sum_{k} f(z_k)=\int_{\Gamma}  \Psi(z)\,f(z)  dz\,.
\end{align}

Now, it may be shown by using Laguerre's differential equation that the function $\Psi(z)$ satisfies
\begin{align}\label{zeros.density.equation}
 \frac{1}{K}\,\Psi'(z)+\frac{c+1/K-z}{z}\,\Psi(z)+\Psi^2(z)+\frac{1}{z}=0\,.
\end{align}
If we employ an asymptotic expansion $\Psi\sim\Psi_0+\frac{1}{K}\Psi_1+\ldots$ for $K\rightarrow \infty$ and retain only the largest order term in $K$, equation \eqref{zeros.density.equation} reduces to
\begin{align}
 (c-z)\,\Psi_0(z)+z\,\Psi_0^2(z)+1=0\,,
\end{align}
which leads to
\begin{align}\label{zeros.density.psi0}
 \Psi_0=\frac{z-c-\sqrt{z^2-2cz-4z+c^2}}{2z}\,.
\end{align}

The equation for the next order in the $1/K$  expansion of $\Psi$ can be straightforwardly obtained and is
\begin{align}\label{zeros.density.psi1}
\Psi_1=-\frac{z\Psi_0'+\Psi_0}{(2z\Psi_0)+c-z}\,.
\end{align}

Finally, the path $\Gamma$ may be continuously shrunk to get an integral over the real axis---considering equations \eqref{zeros.density.psi0} and \eqref{zeros.density.psi1} the asymptotic densities are then given by
\begin{align}
 \label{densidad.D0}D^{(0)}(c,z)&=\begin{cases}
\frac{\sqrt{4z-(z-c)^2}}{2\pi z}\,, &\qquad z_-<z<z_+\\
0\,, &\qquad {\rm otherwise}
                \end{cases}\,,
\\[2mm]
\begin{split}
\label{densidad.D1}D^{(1)}(c,z)&=-\frac{1}{4}\left(\delta(z_--z)+\delta(z_+-z)\right)\\
&\qquad\qquad+\begin{cases}
\frac{1}{2\pi\sqrt{4z-(z-c)^2}}\,, &\qquad z_-<z<z_+\\
0\,, &\qquad {\rm otherwise}
                \end{cases}\,,
\end{split}
\end{align}
where we have defined
\begin{align}
 z_{\pm}=2+c\pm2\sqrt{c+1}\,.
\end{align}

\subsection{The partition function at large $N$}\label{sec.low_temperature}
Let us now apply the asymptotic densities of eigenvalues to the study of the thermodynamics of gases in the NC disc. To that purpose
consider $N$ to be a fixed (large) positive integer, $\ell=0,\pm1,\ldots, \pm(N-1)$ and define consequently $z_k^{\ell}=\lambda_k^\ell/(N-|\ell|)$, where $\lambda_k^\ell$ is the $k$-th zero of the Laguerre polynomial $L_{N-|\ell|}^{|\ell|}$. Then, using the results in Section \ref{appendix.density.zeros}, which can be applied upon the replacements $K\to N-|\ell|$ and $c\to |\ell|/(N-|\ell|)$, the following sum over the roots $z^\ell_k$ for any smooth function $f(z)$ can be approximated as
\begin{align}\label{apo}
  \begin{split}
  \sum_{k=1}^{N-|\ell|} f(z_k^{\ell})
  &=(N-|\ell|)\int_{z_-}^{z_+}\frac{dz}{2\pi z}\ \sqrt{(z_+-z)(z-z_-)}\,f(z)
  +\mbox{}\\[2mm]
  &\mbox{}-\frac{1}{4}\left[f(z_+)+f(z_-)\right]
  +\mbox{}\\[2mm]
  &\mbox{}+\int_{z_-}^{z_+}\frac{dz}{2\pi}\ \frac{1}{\sqrt{(z_+-z)(z-z_-)}}\,f(z)
  +O(N^{-1})\,,
  \end{split}
\end{align}
where $z_\pm$, in the present case, are
\begin{align}
  z_\pm=\left(1\pm\sqrt{\frac{N}{N-|\ell|}}\right)^2\,.
\end{align}

In this way, we study the partition function $\mathcal{Z}_{\theta}$ by using \eqref{apo} to replace the sum over the $k$ index in \eqref{pfgc} by integrals in compact intervals of the real line; this still leaves us with a sum over the $\ell$ index, which will be dealt with  below.

As we will see, the first correction to the partition function due to noncommutativity is $O(\sqrt N)$. Consequently, we will neglect in the following calculation terms in $\log \mathcal{Z}_{\theta}$ which grow slower than $\sqrt N$.

Let us begin by considering the contribution $C_1$ to the partition function of the first line in \eqref{apo},
\begin{align}\label{logZ}
  C_1=\sum_\ell \frac{N-|\ell|}{a}
  \int_{z_-}^{z_+}\frac{dz}{2\pi z}\ \sqrt{(z_+-z)(z-z_-)}
  \ \log\left\{1+ae^{\beta\mu}\,e^{-\frac{1}{\theta}\beta(N-|\ell|) z}\right\}+\ldots
\end{align}
The terms in this sum corresponding to $\ell\neq 0$ can be approximated by means of Euler--Maclaurin formula for any smooth function $F(\ell)$,
\begin{align}\label{em}
  \begin{split}
\sum_{\ell=1}^{N-1}F(\ell)&=\int_1^{N-1}d\ell\ F(\ell)\mbox{}+\frac12\,F(N-1)+\frac12\,F(1)+\mbox{}\\
  &\mbox{}+\sum_{n=1}^\infty \frac{B_{2n}}{(2n)!}\left[F^{(2n-1)}(N-1)-F^{(2n-1)}(1)\right]\,.
  \end{split}
\end{align}

In the case under consideration, the contribution $C_{1,1}$ of the integral term in Euler-Maclaurin formula may be readily obtained after rescaling the integration variables; up to exponentially decreasing contributions it is given by
\begin{multline}\label{z1}
  C_{1,1}=-\frac{N \theta}{a \beta}\,{\rm Li}_2(-ae^{\beta\mu})
  +\frac{\sqrt{N}\theta^\frac32}{a\sqrt\pi\,\beta^{\frac32}}\,{\rm Li}_{\frac52}(-ae^{\beta\mu})+\mbox{}\\+
  \frac{2\sqrt {N}\theta^\frac12}{a\sqrt\pi \beta^{\frac12}}\,{\rm Li}_{\frac{3}{2}}(-ae^{\beta\mu})
  +O(1/\sqrt N)\,,
\end{multline}
where we have introduced the polylogarithm functions
\begin{align}
  {\rm Li}_s(z)=\sum_{n=1}^\infty \frac{z^n}{n^s}\,.
\end{align}
Further details regarding this computation are given in Appendix \ref{integrals}. Moreover, Appendix \ref{boundary} shows that other contributions of order $O(\sqrt N)$ in Euler-Maclaurin approximation come only from the term corresponding to the function $F(\ell)$ evaluated at $\ell=1$. Indeed, as described in detail in Appendix \ref{boundary}, terms in Euler-Maclaurin formula \eqref{em} evaluated at the endpoint $\ell=N-1$ give an exponentially decreasing contribution whereas those corresponding to odd derivatives at $\ell=1$ give contributions that are finite as $N\to\infty$. For this reason, the remaining contribution $C_{1,2}$ from Euler-Maclaurin approximation stems from $F(1)$; a direct computation gives
\begin{align}\label{z2}
\begin{split}C_{1,2}
  &=-\frac{\sqrt {N}\theta^\frac12}{a\sqrt\pi {\beta}^{\frac12}}\,{\rm Li}_{\frac32}(-ae^{{\beta}{\mu}})+\ldots
\end{split}\end{align}

Let us now turn our attention to the term in \eqref{logZ} corresponding to $\ell=0$. In this case, a straightforward computation shows that this contribution is
\begin{align}\label{z3}
  \begin{split}C_{1,3}
  &=-\frac{\sqrt {N}\theta^\frac12}{a\sqrt\pi{\beta}^\frac12}\,{\rm Li}_{\frac32}(-a\,e^{{\beta}{\mu}})
  +O(1/\sqrt N)\,.
\end{split}
\end{align}

The last relevant contribution $C_2$ to the partition function comes from the terms in the second line of \eqref{apo}; this can also be obtained by approximating the sum with the integral term in the Euler-Maclaurin formula,
\begin{align}
\begin{split}  C_2&=-\frac{1}{4a}\ \sum_\ell
  \left(\log\left\{1+ae^{\beta \mu}\,e^{-\frac{1}{\theta}{\beta}(N-|\ell|) z_+}\right\}+
  \log\left\{1+ae^{\beta \mu}\,e^{-\frac{1}{\theta}{\beta}(N-|\ell|) z_-}\right\}\right)+\ldots\,\\\label{z4}
  &=\frac{\sqrt{\pi N}\theta^\frac12}{2a\sqrt{{\beta}}}\,{\rm Li}_{\frac32}(-ae^{{\beta}{\mu}})
  +O(N^0)\,.
\end{split}
\end{align}
In fact, the remaining terms in the Euler-Maclaurin formula can be shown to decrease with large $N$. Similar calculations show that the third line in \eqref{apo} is $O(N^0)$ for large $N$ and hence does not contribute to the leading noncommutative corrections.

We summarize the results of \eqref{z1}, \eqref{z2}, \eqref{z3} and \eqref{z4} in the following  asymptotic expansion for the partition function $\log{\mathcal{Z}}_{\theta}$ per unit area:
\begin{align}\label{pf}
  \frac{a}{\pi R^2}\log{\mathcal{Z}_{\theta}}\simeq -\frac{1}{4\pi\beta}\,{\rm Li}_2(-ae^{\beta\mu})
  +\frac{1}{4\pi^{\frac12}\beta^{\frac12}R}\,{\rm Li}_{\frac32}(-ae^{\beta\mu})
  +\frac{\theta}{2\pi^{\frac32}\beta^{\frac32}R}\,{\rm Li}_{\frac52}(-ae^{\beta\mu})\,.
\end{align}
Note that we have used $4\pi\theta N\sim\pi R^2$.

Some remarks are now in order. First of all, expression \eqref{pf} is well-behaved as $\theta$ runs to zero once we have fixed the radius of the disc. In fact, its leading term corresponds to the two-dimensional gas in the commutative plane: for a free non-relativistic particle in the whole $\mathbb{R}^2$ direct integration in phase space gives
\begin{align}
    \log \mathcal{Z}_{\,\mathbb{R}^2}=\frac{1}{a}
    \ {\rm Tr}\ \log\left\{1+a\,e^{-\beta(-\triangle-\mu)}\right\}
    =-\frac{1}{a}\,{\rm Vol}(\mathbb{R}^2)\,\frac{1}{4\pi\beta}\ {\rm Li}_2(-ae^{\beta\mu})\,.
\end{align}
Note however that the second term in \eqref{pf} does also remain, as a boundary contribution, in the limit $\theta\to 0$. In fact, an additional check of the first two terms in the r.h.s.\ of \eqref{pf} is obtained from the partition function $\mathcal{Z}_0$ in the ordinary commutative disc of radius $R$,
\begin{align}
\begin{split}
\frac{a}{\pi R^2}\log \mathcal{Z}_0 &=\frac{1}{a}\ \sum_{\ell\in\mathbb{Z}}\ \sum_{k=1}^{\infty}
  \ \log\left\{1+a e^{-\beta\left(\frac{(j_k^\ell)^2}{R^2}-\mu\right)}\right\}\,\\
    &= -\frac{1}{4\pi\beta}\,{\rm Li}_2(-ae^{\beta\mu})
  +\frac{1}{4\pi^\frac12 \beta^\frac12 R}\,{\rm Li}_\frac32(-ae^{\beta\mu})+\ldots\,,
\end{split}
\end{align}
where $j_k^\ell$ denotes the $k$-th zero of the Bessel function $J_\ell(\cdot)$. This result can be obtained with standard heat-kernel methods.

\begin{figure}[t]
\begin{minipage}{.48\textwidth}
\captionsetup{width=.8\textwidth}
\includegraphics[width=.9\textwidth]{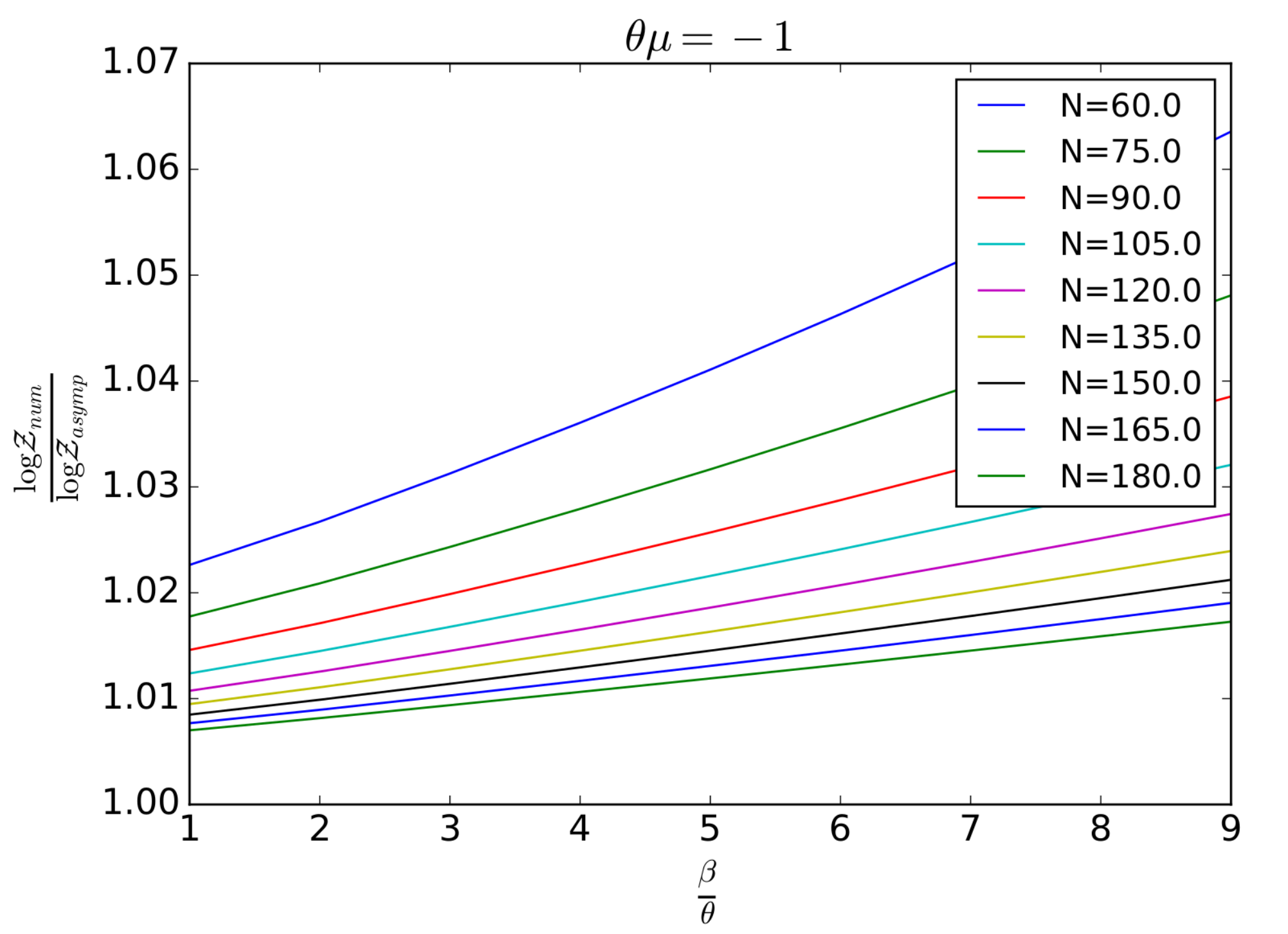}
\caption{\tiny Quotient $\frac{\log \mathcal{Z}_{\rm num}}{\log \mathcal{Z}_{\rm asymp}}$ as a function of $\frac{\beta}{\theta}$, the inverse temperature in units of $\theta$, for $\theta\mu=-1$ and several values of $N=60,\, 75\, \ldots 180$. The value of $N$ increases from the upper to the lower curve in the plot (in colors in the original), showing the accuracy of our analytic asymptotic expansion.\label{fig.largeN}}
\end{minipage}
\begin{minipage}{.48\textwidth}
\captionsetup{width=.8\textwidth}
\includegraphics[width=.9\textwidth]{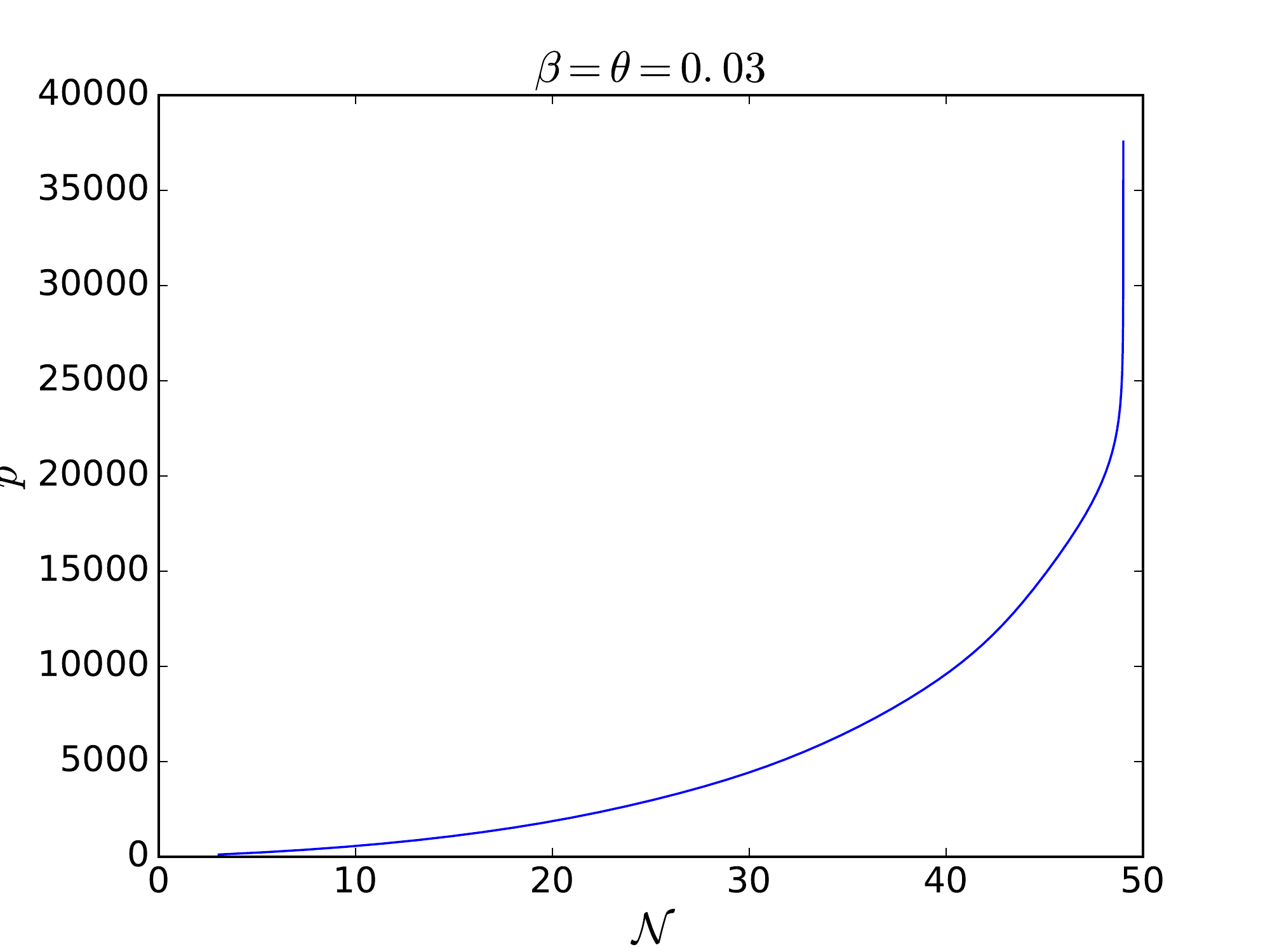}
\caption{\tiny Pressure $p$ of a fermionic system as a function of the number of particles $\mathcal{N}$ for $\beta=\theta=0.03$ and $R=1$. As expected, the pressure diverges as the number of particles tends to its (finite) maximum value.\vspace{1.cm}\label{fig.degenerate_fermi}}
\end{minipage}
\end{figure}

Secondly, let us mention that several numerical tests support the validity of asymptotic expansion \eqref{pf}. As an example, we display in Figure \ref{fig.largeN} the quotient $\log{\mathcal{Z}_{\rm num}}/\log{\mathcal{Z}_{\rm asymp}}$ of the logarithm of the partition function, $\log{\mathcal{Z}_{\rm num}}$, computed through a numerical evaluation of the Laguerre zeros in \eqref{pfgc} and $\log{\mathcal{Z}_{\rm asymp}}$, its large-$N$ expansion \eqref{pf}, as a function of the rescaled inverse temperature $\frac{\beta}{\theta}$ and for several values of $N$. This figure, corresponding to fermions ($a=1$) at a nonvanishing chemical potential ($\theta\mu=-1$), shows that both results agree in the large-$N$ limit.

Thirdly, and related to the previous comment, when one considers fermionic or Maxwell-Boltzmann statistics,  \eqref{pf} additionally shows that the large volume expansion we have obtained breaks down whenever we consider a large chemical potential such that $\theta \mu \sim N$, viz.\ when we consider a system with large number of particles---it is so simply because polylogarithm functions behave as ${\rm Li}_s(-z)\sim (\log z)^s$ for large and real\footnote{For the Maxwell-Boltzmann gas the polylogarithms actually reduce to linear functions, viz. ${\rm Li}_s(-z)\sim z$. However the conclusions do not differ.} $z$.

Finally, we would like to insist on the relevance of expansion \eqref{pf} as a report of the leading noncommutative corrections to a quantum problem which, being confined to a compact region, can be regarded as a more realistic approach to a physical setting. As we have mentioned in the Introduction, there are many studies devoted to the analysis of noncommutative corrections as perturbations of the Hamiltonian based on the leading terms in the Moyal product. Of course, this analysis does not capture the full non-local character of Moyal interactions. On the other hand, several noncommutative systems are known that admit explicit solutions which, however, are non-perturbative in $\theta$. This is a well-known fact in many noncommutative theories, for which the limit $\theta\to 0$ is not smooth due, precisely, to the non-locality of the interactions and the consequent intertwining between the UV and the IR regimes. The problem considered in this article makes it possible to obtain a perturbative analysis of noncommutativity from the full noncommutative solution.

Before considering the high-temperature regime, we discuss certain aspects of the results we have obtained applied to the different statistics.

\subsection{Maxwell-Boltzmann statistics}
Let us first analize the large-$N$ behaviour of a Maxwell-Boltzmann gas. If we consider a finite number of states, the limit $\beta\rightarrow\infty$ of eq.\ \eqref{pfgc} shows that only the lowest energy level should be populated. In our case, as the volume gets larger the lowest energy state tends to zero. This is consistent with the following behaviour of the energy density $e=\frac{E}{\pi R^2}$, in terms of the particle density $n=\frac{\mathcal{N}}{\pi R^2}$, for low temperatures:
\begin{align}
 e\simeq n\left(\frac{1}{\beta}
 +\frac{\sqrt\pi}{2\sqrt\beta R}
 -\frac{\theta}{\sqrt\pi \beta^{\frac32} R}\right)\,,
\end{align}
which shows that corrections to the equipartition principle also receive contributions from noncommutativity.

Additionally the entropy density admits a kind of Sackur-Tetrode expression,
\begin{align}
    s\simeq n\left(2-\log(4\beta n\pi)
    -\frac{\sqrt{\pi\beta}}{2R}
    -\frac{3\theta}{\sqrt{\pi\beta} R}\right)\,.
\end{align}
The divergence of this expression for low temperatures $\beta^{-1}\ll n$ points out the non-physicality of the model in this regime.

\subsection{Fermi-Dirac statistics}

The fermionic character of a commutative Fermi gas can be understood from many thermodynamical quantities, among them the entropy density $s$ and the heat capacity per particle $\frac{C_V}{\mathcal{N}}$, which can be readily derived from \eqref{pf}. In particular, in the regime  $N\gg \beta\mu\gg1$ one obtains,
\begin{align}
\begin{split}\label{entropy.c_v.largeN.f}
    s&\simeq \frac{1}{\beta}\left(\frac{\pi}{12}-\frac{\pi}{12R\sqrt{\mu}}-\frac{\theta \sqrt\mu}{3R}\right)\,,\\
    \frac{C_V}{\mathcal{N}}&\simeq \frac{\pi^2}{3\beta \epsilon_F}
    \left(1-\frac{1}{R\sqrt{\epsilon_F}}-\frac{4\sqrt{\beta\epsilon_F}\theta}{\pi R}\right),
\end{split}
\end{align}
where we have introduced the Fermi energy $\epsilon_F$. Other quantities such as the  energy density $e$, the density of particles $n$ and the full expression of $s$ and $C_V$ in terms of $\beta\mu$ are left to Appendix \ref{appendix.fermion}.

On the one hand, from eq.\ \eqref{entropy.c_v.largeN.f} for the entropy it can be seen that Nernst theorem, viz.\ the third law of thermodynamics, is satisfied by this noncommutative gas. The same conclusion can be reached employing the full expression \eqref{fermion.s-largeN3} in Appendix \ref{appendix.fermion}.

On the other hand, as is shown in \eqref{entropy.c_v.largeN.f}, the heat capacity of this gas at constant volume behaves as $\beta^{-1}$ for low temperatures, even considering the noncommutative corrections. This can be understood in the usual way: at low temperatures only the fermions near the Fermi energy can be excited, so that the system increases its energy by a factor $T^2$ and $C_V$ is therefore proportional to $T$.

\subsection{Bose-Einstein statistics}

An interesting effect which can arise in a bosonic system is Bose-Einstein Condensation (BEC). This effect is usually understood \cite{Pathria} by considering the behaviour of the density $n$ of particles, given by eq (\ref{termodinamica}), as
\begin{align}\label{BECdensity}
 n&:= \frac{1}{\pi R^2}\,\frac{1}{e^{\beta(\lambda^0_1-\mu)}-1}+n_e\,,
\end{align}
in terms of the contributions from the fundamental energy level and the excited ones, $n_e$.

For a system of a given density, one expects that in the thermodynamical limit, i.e., $\pi R^2\rightarrow\infty$, the infinite number of particles will be more or less evenly distributed over the infinite number of levels, so that the density of a single state, like the ground state, should be zero. However, for some systems there exists an upper bound $n_c(\beta)$ on the density of the excited states, called critical density. As a consequence, in order to have a system with a density higher than the critical one, there should be a non-zero contribution from the fundamental level, as can be seen from eq (\ref{BECdensity}). In those cases it is said that we are in the presence of a BEC. However, this is not the case under study.

Indeed, let us consider the expression for the density of particles computed from the partition function \eqref{pf},
\begin{align}
\label{boson.n-largeN}n\simeq \frac{\log(1 - e^{\beta \mu})}{4\pi \beta} + \frac{1}{4\pi^{1/2}}\frac{1}{  \beta^{1/2} R} \text{Li}_{1/2}(e^{\beta\mu}) +\frac{1}{2\pi^{3/2}} \frac{\theta }{\beta R} \text{Li}_{3/2}(e^{\beta\mu})\,.
\end{align}
It is clear from this expression that the density can be increased indefinetely by choosing a chemical potential $\mu$ as close to zero as needed, since both the first and second term in the r.h.s\ of \eqref{boson.n-largeN} are divergent in this regime. In other words, the density of states near the ground state is such that they are able to contain an indefinitely large number of particles.  Although it is not a surprise---since in the large volume limit one expects the noncommutative theory to resemble the commutative one---we have not previously seen a proof of this. In Appendix \ref{appendix.nobec} we sketch another proof which does not involve the use of Laguerre's density of zeros.

\section{The high-temperature regime}\label{s-high}

In this section, we provide an analysis of the high temperature regime of a gas in the NC disc. This regime should be understood as that in which $\frac{\beta}{\theta} \lambda_k^\ell \ll 1$, while the fugacity $z=e^{\beta\mu}$ remains fixed. Since the largest Laguerre zero scales as $N^2$, this condition means $\beta\ll \theta^3/R^4$. As a consequence, the subsequent expressions do not admit the commutative limit $\theta\to 0$. In this regime, the partition function $\mathcal{Z}_{\theta}$ of eq. \eqref{pfgc} admits the following asymptotic expansion:
\begin{equation}
  \log \mathcal{Z}_{\theta}=\sum_{\ell,k} \frac{\log(1+az)}{a}-\frac{z}{1+az}\frac{\beta}{\theta}\sum_{\ell,k}\lambda^\ell_k+\frac{1}{2}\frac{z}{(1+a z)^2}\frac{\beta^2}{\theta^2}\sum_{\ell,k}\left(\lambda^\ell_k\right)^2
  +\ldots
\end{equation}
It is a straightforward exercise to show that sums of powers of the roots of a polynomial can be computed from its coefficients. In the case under consideration, i.e., zeros $\lambda^\ell_k$ of the Laguerre polynomials $L^{|\ell|}_{N-|\ell|}(\lambda)$, the known expressions for their coefficients \cite{A-S} lead to
\begin{align}
\begin{split}
  \sum_{\ell,k}\lambda^\ell_k&=N^3\,\\
  \sum_{\ell,k}\left(\lambda^\ell_k\right)^2&=
  \frac13N^2(5N^2-3N+1)\,.\label{slc}
\end{split}
\end{align}
From these expressions we get the following high-temperature behaviour for the grand canonical partition function:
\begin{align}\label{bosons.partition}
\begin{split}\log \mathcal{Z}_{\theta} &=\frac{N^2}{a}\log(1+az)+ N^2\frac{z}{1+az}\frac{N}{\theta} \beta +\frac{N^2(5N^2-3N+1)}{6(1+az)^2} z\, \frac{\beta^2}{\theta^2}+\cdots
\end{split}
\end{align}
Notice that this expansion is not compatible with a large-$N$ expansion, i.e., a large volume expansion. 

Performing a straightforward computation, we obtain from eqs.\ \eqref{termodinamica} and \eqref{bosons.partition} the following expressions for the number of particles, mean energy and entropy:
\begin{align}\label{equations.statistics}
 \begin{split}
  \mathcal{N}&=\frac{N^2 z}{1+a\,z}-\frac{N^3 z}{(1+a\,z)^2} \frac{\beta}{\theta}
  + \frac{N^2(1-3N+5N^2)}{6} \frac{(1-a\,z)z}{(1+a\,z)^3} \frac{\beta^2}{\theta^2}+\cdots\\[2mm]
  E&=N^3\frac{z}{1+az}\frac{1}{\theta}-\frac{N^2(5N^2-3N+1)}{3(1+az)^2} z\, \frac{\beta}{\theta^2}+\cdots \\[2mm]
  S&=\frac{\log{(1+az)}}{a}\, N^2+\mbox{}\\
    &-\frac{N^2 z}{1+az}\log z+\frac{N^3 z}{1+az}\log z \frac{\beta}{\theta} -\frac{N^2(5N^2-3N+1)}{6} \frac{z}{(1+a\,z)}\frac{\beta^2}{\theta^2}+\ldots
 \end{split}
\end{align}
It is clear that as the temperature increases the mean energy per particle tends to the average of the energy levels---independently of the number of particles and the statistics of the system---as expected for a system with a finite number of levels. This is just a consequence of the fact that the particles tend to be uniformly distributed in this regime. To the same conclusion points out the statistical interpretation of the entropy that can be derived from \eqref{equations.statistics} at infinite temperature. Needless to say, we have the usual restriction that the number of particles is bounded for a fermionic gas, but unbounded for a Maxwell-Boltzmann and Bose-Einstein gas. As a consequence, for a fermionic gas, as the number of particles reaches its maximum, $N^2$, the fugacity $z$ tends to infinity and the pressure of the system diverges. This dependence of the pressure for a fermionic gas as a function of the number of particles has been plotted in Figure \ref{fig.degenerate_fermi}, for $\beta=\theta=0.03$ in units of $R^2=1$.

One interesting measurable quantity is the heat capacity at constant volume for a small number of particles\footnote{Interestingly, the suitable parameter to perform a ``low density'' expansion is in this case $\mathcal{N}\ll 1$, in contrast to what happens in the commutative case, in which its role is played by the product of the density with the cube of the thermal length.} $\mathcal N\ll 1$, which shows a $\beta^2$ behaviour together with a factor that depends on the statistics of the system,
\begin{align}
 C_{V}= \frac{1}{3}  ( N-1) ( 2 N -1) \left(1-a \frac{\mathcal{N}}{N^2}\right) \mathcal{N}\frac{\beta^2}{\theta^2}+\ldots
\end{align}
Interestingly the statistic's correction to the heat capacity depends on $\frac{\mathcal{N}}{N^2}$ the density of particles per volume, instead of the density of particles. Since the system has a finite number of available states, $C_V\to 0$ in the limit $\beta\to 0$.

Finally, it can be shown that the gas satisfies an equation of state that resembles that of a classical gas when one considers the {\rm dilute} limit,
\begin{align}
\beta\,p\,\pi R^2 = \mathcal{N}\left[1+ a\mathcal{N} \left(\frac{1}{2N^2}+\beta^2 \left(\frac{1}{3\theta^2}+\frac{1}{6 N^2\theta^2}-\frac{1}{2N\theta^2}\right)\right) \right].
\end{align}
In the fermionic and bosonic cases there exist quantum corrections which can be explained in terms of an effective repulsion and attraction, respectively. In any case, this expression differs from the one obtained in the commutative case: the corrections, instead of being proportional to the density of particles and a power of the temperature, are proportional to the number of particles and a polynomial of the inverse temperature and volume.

\section{Conclusions}\label{s5}

We have studied the thermodynamics of a quantum gas confined in a bounded region of the Moyal plane. We have considered Fermi-Dirac, Bose-Einstein and classical gases.

On the one hand, the use of the density of Laguerre zeros leads to an expression for the grand canonical partition function which isolates the leading noncommutative effects. Thus, as a first result, we show how to obtain this density of zeros order by order in terms of the effective size $N$ of the NC disc. Once we have this expression at our disposal, we are able to compute the energy and the entropy for a Maxwell-Boltzmann gas, showing that at the NC level the equipartition of energy is broken; also a generalized Sackur-Tetrode expression is derived.

For fermions, we have shown that essential fermionic aspects at low temperatures are not modified by noncommutativity. In fact, at zero temperature the system is in a state for which all the energy levels below the Fermi energy are filled. Moreover, the excitations at finite temperature are such that the heat capacity is proportional to $\beta$.

We have also proved that there is no Bose-Einstein condensation for a non-relativistic gas in the noncommutative case. This is not a surprise, since this mimics the behaviour of a gas in the commutative case. However, it is interesting to notice that the first noncommutative correction can only be populated by a limited number of particles in the limit of interest for the Bose-Einstein condensation.

On the other hand, in the high-temperature regime we have made use of the algebraic properties of the roots of the Laguerre polynomials to obtain an expansion of the grand canonical partition function. In the diluted limit the corrections to the equation of state of a classical gas are obtained. Most notably, the leading noncommutative corrections reinforce the statistical corrections, i.e., act as an effective repulsion for fermions and an attraction for bosons. They show however an unusual behaviour in terms of the volume of the system, so that they are not simple functions of the density of particles. Additionally, the behaviour of the heat capacity is dictated by an uncommon quadratic law at large temperatures.

\bigskip

\noindent\textbf{Acknowledgements}:  This work was partially supported by grants from CONICET (PIP 01787), ANPCyT (PICT-2011-0605) and UNLP (Proy.~11/X615), Argentina. SF acknowledges support from the Physics Department of the UNLP and the DAAD under the ALE-ARG program. The work of PP is supported by CONICET.

\appendix

\section{Contributions of \eqref{logZ} for $l\neq0$ via the Euler-Maclaurin formula} \label{contributions_euler_maclaurin}
In this appendix we show how to compute the main contributions to $\eqref{logZ}$ in an asymptotic expansion in $N$.  We will consider positive angular momentums $\ell$, since the results are the same for negative values.
\subsection{The integral term } \label{integrals}
This term may be written as
\begin{align}\label{em-lead}
  \begin{split}C_{1,1}&= 2\int_1^{N-1}d\ell\ \frac{N-\ell}{a}\
  \int_{z_-}^{z_+}\frac{dz}{2\pi z}\ \sqrt{(z_+-z)(z-z_-)}\times\mbox{}\\[2mm]
  &\mbox{}\times\log\left\{1+ae^{{\beta}{\mu}}\,e^{-\frac{1}{\theta}{\beta}(N-\ell)z}\right\}+\ldots .
  \end{split}
\end{align}
After rescaling $(1-\ell/N)\,z\to z$, and replacing $\ell\to N(1-\ell)$, we obtain
\begin{align}\label{int-em-1st}
\begin{split}
    C_{1,1}&=-\frac{N^2}{\pi a}\sum_{n=1}^\infty \frac{(-ae^{{\beta}{\mu}})^n}{n}\times\mbox{}\\[2mm]
    &\mbox{}\times \int_{0}^{1-\frac1N}d\ell
    \ \int_{(1-\sqrt{\ell})^2}^{(1+\sqrt{\ell})^2}dz\ \frac{e^{-\frac{1}{\theta}n{\beta} N z}}{z}
    \ \sqrt{\left[(1+\sqrt{\ell})^2-z\right]\left[z-(1-\sqrt{\ell})^2\right]}+\ldots
\end{split}
\end{align}
The double integral in \eqref{int-em-1st} can be  explicitly computed after inverting the order of integration and leads to \eqref{z1}. However, it should be noticed that in \eqref{int-em-1st} we have replaced the integration domain $\ell\in(\frac1N,1-\frac1N)$ by the interval $\ell\in(0,1-\frac1N)$ because integration in $\ell\in(0,\frac1N)$ is exponentially decreasing as $N\to \infty$. Indeed, with the same procedure used to obtain eq.\ \eqref{int-em-1st} one can show that the absolute value of the integration in the interval $(0,\frac1N)$ is bounded by
\begin{multline}
  \frac{4N^2}{\pi |a|}\sum_{n=1}^\infty \frac{(| a| e^{\beta\mu})^n}{n}
  \int_{1-\frac{1}{\sqrt{N}}}^{1+\frac{1}{\sqrt{N}}}dz\ e^{-\frac{1}{\theta}n\beta N z}
  \int_{\frac{z+1-\frac1N}{2\sqrt{z}}}^{1}d\ell\ \sqrt{1-\ell^2}\,\\
  \leq \frac{4\theta}{3\pi\beta a}N^{-\frac12}\sum_{n=1}^\infty \frac{( | a | e^{\beta\mu})^n}{n^2}
  \,e^{-\frac{1}{\theta}n\beta N}\,\sinh{(n\beta\sqrt{N}/\theta)}.
\end{multline}
Thus, the omitted integration at the lower boundary in \eqref{int-em-1st} does not contribute to the asymptotic expansion for large $N$.

\subsection{The boundary terms} \label{boundary}

It will be proved in this section that the only missing contribution to \eqref{logZ} using the Euler-McLaurin formula comes from $F(1)$. In order to do that, we study the values of the function
\begin{align}
  F(\ell)=\frac{N}{2\pi a}\int_{w_-}^{w_+}\frac{dz}{z}\ \sqrt{(w_+-z)(z-w_-)}
  \ \log\left\{1+ae^{\beta\mu}\,e^{-\frac{1}{\theta}\beta N z}\right\}
\end{align}
and its derivatives, at $\ell=1$ and $\ell=N-1$, as $N\to\infty$. The endpoints take the values
\begin{align}
  w_\pm=\left(1\pm \sqrt{1-\frac{\ell}{N}}\right)^2\,,
\end{align}
so that, for large $N$, the integration is performed on a small interval around $z=1$, if $\ell=N-1$, or on an interval close to the segment $z\in[0,4]$, if $\ell=1$. In any case, the endpoints satisfy $0<w_-<1<w_+<4$.

To simplify the proof, we expand the logarithm and we recast the function $F(\ell)$ as
\begin{align}
  F(\ell)=-\frac{N}{2\pi a}\ \sum_{n=1}^\infty\ \frac{(-ae^{\beta\mu})^n}{n}
  \left\{F_n^+(\ell)-F_n^-(\ell)\right\}
\end{align}
using the following definitions:
\begin{align}
\begin{split}
F_n^\pm(\ell)&=\int_{1}^{w_\pm}dz\ \sqrt{\pm w_\pm\mp z}\ h_n^\pm(z)\,,\\
h_n^\pm&=\sqrt{\pm z\mp w_\mp}\ \frac{e^{-\frac{1}{\theta}n\beta N z}}{z}\,.
  \end{split}
\end{align}
Since, in our case, the dependence of $F$ on $\ell$ is only through $w_{\pm}$, it is clear that we may as well just study the derivatives of $F$ with respect to $w_{\pm}$. Consider for example the first derivative of $F_n^\pm$ with respect to $w_\pm$:
\begin{align}\label{b.derivative}
  \partial_{w_\pm}F_n^\pm&=\int_{1}^{w_\pm}dz
  \ \left(-\partial_z\sqrt{\pm w_\pm \mp z}\right)\ h_n^\pm(z)\nonumber\\[2mm]
  &=\sqrt{\pm w_\pm\mp1}\ h_n^\pm(1)+\int_{1}^{w_\pm}dz\ \sqrt{\pm w_\pm\mp z}\ \partial_zh_n^\pm(z)\,,
\end{align}
where we have recasted the derivative with respect to $w_{\pm}$ as a derivative with respect to $z$ and integrated by parts. Similarly, one obtains
\begin{align}\label{b.2derivative}
  \partial^2_{w_\pm}F_n^\pm=\frac{h_n^\pm(1)}{2\sqrt{\pm w_\pm\mp1}}
  \pm\sqrt{\pm w_\pm\mp1}\ \partial_z h_n^\pm(1)
  \pm\int_{1}^{w_\pm}dz\ \sqrt{\pm w_\pm\mp z}\ \partial^2_zh_n^\pm(z)\,,
\end{align}
and analogous expressions for higher derivatives. Note that for $\ell=N-1$ the boundary terms that appear after successive integration by parts contain a divergent part as $N\to\infty$ because $w_\pm\simeq 1\pm\frac{2}{\sqrt N}$. Moreover, there are additional divergent factors coming from the chain rule:
\begin{align}\label{der}
  \partial_\ell=-\frac{1}{N}\,\frac{\sqrt{w_+}}{\sqrt{w_+}-1}\,\partial_{w_+}
  +\frac{1}{N}\,\frac{\sqrt{w_-}}{1-\sqrt{w_-}}\,\partial_{w_-}\,.
\end{align}
However, these divergencies are all cancelled by the exponential decrease of $h_n^\pm(z)$ and its derivatives at $z=1$.

As regards the remaining integral in \eqref{b.derivative} and \eqref{b.2derivative}, the higher derivatives of $h_n^+$ (respectively, $h_n^-$) only diverge at $z=x_-$ (respectively, $x_+$), which does not belong to the integration domain. For this reason, all boundary terms in Euler-Maclaurin formula corresponding to $\ell=N-1$ are exponentially suppressed as $N$ grows.

The boundary terms at $\ell=1$ deserve further attention because the functions $h_n^-(z)$ and its derivatives contain increasing powers of $1/z$ which give divergent contributions to $F^-_n$ and its derivatives when integrated around $w_-$, since $w_-$ tends to zero as $N\to\infty$. Nevertheless, we will now show that the integrals
\begin{align}
  \int_{w_-}^{1}dz\ \sqrt{z-w_-}\ \partial^k_zh_n^-(z)
\end{align}
are bounded, for any $k\in\mathbb{Z}^+$, as $N\to\infty$. The $k$-th derivative of $h_n^-$ contains a term proportional to $N^k$ coming from the derivatives of the exponential, but also a term proportional to $1/z^{1+k}$ coming from the derivatives of the denominator $1/z$; in fact, the latter dominates for large $N$. Indeed,
\begin{align}
    \int_{w_-}^{1}dz\ \sqrt{z-w_-}\sqrt{w_+-z}\ \frac{e^{-\frac{1}{\theta}n\beta N z}}{z^{1+k}}
    \sim\int_{\frac{1}{4N^2}}^{1}dz\ \sqrt{4-z}\ \frac{e^{-\frac{1}{\theta}n\beta N z}}{z^{k+\frac12}}
    \sim N^{2k-1}\,.
\end{align}
Therefore, each derivative $\partial_{w_-}$ in \eqref{der} acting on $F_n^-$ adds a factor $N^2$. Note also that, for large $N$,
\begin{align}\label{factor}
  \frac{\sqrt{w_-}}{1-\sqrt{w_-}}\sim \frac1N\,,\qquad
  \partial_{w_-}\left(\frac{\sqrt{w_-}}{1-\sqrt{w_-}}\right)\sim N\,,
\end{align}
and so forth, so that each derivative $\partial_{w_-}$ adds a factor $N^2$ independently on the function it acts on--- consequently, the $k$-th derivative of $F^-_n$ contains a divergence $N^{2k-1}$. However, each derivative also contains a factor $1/N$, explicitly written in \eqref{der}, and a second factor $1/N$ arising from the term in \eqref{factor}, which also appears in \eqref{der}. These factors cancel the divergence $N^{2k}$ and the $k$-th derivative of $F^-_n$ behaves as $1/N$. Since its contribution to $\log{\mathcal{Z}_\theta}$ contains an additional factor $N- | \ell |$ (see \eqref{logZ}) then all boundary terms corresponding to derivatives at $\ell=1$ in Euler-Maclaurin formula are bounded as $N\to\infty$. Differently, the term with no derivative at $\ell=1$ grows as $\sqrt{N}$ as has already been computed in \eqref{z2}.

\section{Other fermionic thermodynamic quantities for small $\theta$}\label{appendix.fermion}
In this appendix we show some expressions for the low-temperature regime of a Fermi gas. In particular, one can compute the particle density $n$,
\begin{align}\label{fermion.s-largeN1}
    n&= \frac{\log(1 + e^{\beta \mu})}{4\pi \beta}
    +\frac{1}{4\pi^{1/2}}\frac{1}{  \beta^{1/2} R} \text{Li}_{1/2}(-e^{\beta\mu})
    +\frac{1}{2\pi^{3/2}} \frac{\theta }{\beta R} \text{Li}_{3/2}(-e^{\beta\mu})\,,
\end{align}
the energy density $e$,
\begin{align}\label{fermion.s-largeN2}
    e&=-\frac{1}{4 \beta^2 \pi } {\rm Li}_{2}(-e^{\beta\mu})
    +\frac{1}{8\pi^{1/2} \beta^{3/2} R}\text{Li}_{3/2}(-e^{\beta\mu})
    +\frac{3 }{4\pi^{3/2}} \frac{\theta }{\beta^{3/2}  R}\text{Li}_{5/2}(-e^{\beta\mu})\,,
 \end{align}
the entropy density $s$,
\begin{align}\label{fermion.s-largeN3}
    s&=-\frac{\mu } {4 \pi } \log( 1 + e^{\beta\mu})- \frac{1 }{4 \pi^{1/2}}\frac{\beta^{1/2} \mu }{ R}
    {\rm Li}_{1/2}(-e^{\beta\mu})\\
    &+ \frac{ (3 \pi - 4 \mu \theta) }{8 \pi^{3/2}}\frac{1}{ \beta^{1/2} R} {\rm Li}_{3/2}(-e^{\beta\mu})
    -\frac{1 }{2 \pi \beta}  {\rm Li}_{2}(-e^{\beta\mu})
    +\frac{5 }{4 \pi^{3/2}}\frac{\theta }{ \beta^{3/2} R}{\rm Li}_{5/2}(-e^{\beta\mu})\,,\nonumber
\end{align}
and the heat capacity per unit volume,
\begin{align}
  {c_V}&= -\frac{2 \text{Li}_2(-e^{\beta\mu})+\log^2(1+e^{\beta\mu})
  + e^{-\beta\mu} \log^2(1+e^{\beta\mu})   }{4\pi\beta}\\
  &+\frac{1}{32} \frac{1}{\sqrt{N}}\frac{1}{\pi^{3/2}\beta^{3/2} \theta^{1/2}}
  \Bigg\lbrace 4\pi\beta(1+e^{-\beta\mu})^2 \log^2(1+e^{\beta\mu}) \text{Li}_{-1/2}(-e^{-\beta\mu})\nonumber\\
  &+4(1+e^{-\beta\mu})\log(1+e^{\beta\mu} ) (-\pi \beta+2(1+e^{-\beta\mu}) \theta
  \log(1+e^{\beta\mu})) \text{Li}_{1/2}(-e^{\beta\mu})\nonumber\\
  &+ 3 (\pi \beta- 8 \theta (1+e^{-\beta\mu}) \log(1+e^{\beta\mu})) \text{Li}_{3/2}(-e^{\beta\mu})
  + 30 \theta \text{Li}_{5/2}(-e^{\beta\mu}) \Bigg\rbrace\nonumber\,.
 \end{align}

\section{Alternative proof for the nonexistence of BEC for a gas in the NC disc}\label{appendix.nobec}

In this appendix we will give another proof of the fact that BEC does not take place for a  gas in the NC disc. The following proof consists in showing that for a chemical potential conveniently chosen, the density $n_e$ of the excited states can take a value as large as we want.

To begin, it should be noted that according to eq. (\ref{Ndef}) and once we have fixed the NC parameter $\theta$, taking large volumes implies large $N$. We will also need the following results regarding the zeros $\alpha_k^{\ell}$ of the Laguerre polynomials $L_{N}^{\ell}$, and the zeros $j_k^{\ell}$ of Bessel functions \cite{calogero:1978} and \cite{A-S}

\begin{align}
 \label{sum_besel}\sum_{k=1}^{\infty}(j^\ell_k)^{-2}&=\frac{1}{4(\ell+1)}\\
 \label{L-expansion} \alpha^\ell_k&=\frac{\left(j^\ell_k\right)^2}{4N-2\ell+2}+O(N^{-3})\,,
\end{align}
where $\ell$ and $k$ are supposed to be fixed in order to the asymptotic expansion (\ref{L-expansion}) to be valid.

By virtue of eq. (\ref{sum_besel}) and the divergence of the series of $\ell^{-1}$, we can choose $\ell_0$ and $k_0(\ell)$ such that for an arbitrary large number M we have
\begin{align}
 \label{series_l}\sum_{\ell=0}^{\ell_0} \frac{1}{4(\ell+1)}&> \theta(M+1)\\
 \label{series_k}\sum_{k=1}^{k_0(\ell)}(j^\ell_k)^{-2}&>\frac{1}{4(\ell+1)}-\frac{1}{\ell_0}.
\end{align}
Then it is straightforward to see, using eqs. (\ref{L-expansion}-\ref{series_k}) and choosing N large enough, that the density of the excited levels will be larger than the sum of the first terms bounded by $\ell_0$ and $k_0(\ell)$. Thus, our proposition follows:
 \begin{align}
  \begin{split}
  n_e\left(\mu=\frac{\alpha^0_1}{\theta}\right)
  &>\frac{1}{4\theta N}\sum_{\ell=0}^{\ell_0}\sum_{k=0}^{k_0(\ell)}
  \frac{1}{e^{\beta(\alpha^\ell_k-\alpha^0_1)/\theta}-1}\\
  &=\frac{1}{\beta}\sum_{\ell=0}^{\ell_0}\sum_{k=0}^{k_0(\ell)} \frac{1}{(j^\ell_k)^2-(j^0_1)^2}(1+O(N^{-2}))\\
  &\geq\frac{1}{\theta}\sum_{\ell=0}^{\ell_0}\sum_{k=0}^{k_0(\ell)} \frac{1}{(j^\ell_k)^2}(1+O(N^{-2}))\\
  &\geq M +O(N^{-2}).
\end{split}
 \end{align}

The proof for the case of the $\lambda_k^{\ell}$ zeros of the Laguerre $L_{N-\ell}^{\ell}$ should be clear for the reader. Indeed since the energy levels we consider have a bounded angular momenta, the $\lambda_k^{\ell}$ zeros satisfy a relation analogue to \eqref{L-expansion}.

It is worth noting that the key in the preceding proof was to realize that the thermodynamical limit is analogous to the commutative one, in the sense that both are given by the $N\rightarrow \infty$ limit. Since it is known that the 2D commutative system of bosons admits no BEC, it could had been expected that the same result would had proven true in the NC system.




\end{document}